\documentclass[12pt]{iopart}
\usepackage{setstack}
\usepackage{cite}
\usepackage[pdftex]{graphicx}
\usepackage{latexsym}
\usepackage{multirow}
\usepackage{dcolumn}
\usepackage{subfigure}
\newcommand{\Sterm}{{\rm S}}
\newcommand{\Pterm}{{\rm P}}

\newcommand{\Ga}{{\rm Ga}}
\newcommand{\G}{{\rm G}}
\newcommand{\g}{\rm g}
\newcommand{\timeT}{\rm t}
\newcommand{\halfh}{\textstyle{1 \over 2}h}

\newcommand{\vek}[1]{\mbox{\boldmath $  #1$}}
\newcommand{\be}{\begin{equation}}
\newcommand{\ee}{\end{equation}}

\begin{document}


\title[Doppler cooling of gallium 2]{Doppler cooling of gallium atoms: 2. Simulation in complex multilevel systems}
\author{L Rutherford$^1$, I C Lane$^2$, J F McCann$^1$}
\address{$^1$Department of Applied Mathematics and Theoretical Physics, Queen's University, Belfast BT7 1NN, UK} 
\address{$^2$Innovative Molecular Materials Group, School of Chemistry and Chemical Engineering, Queen's University, Belfast BT9 5AG, UK}
\ead{i.lane@qub.ac.uk}

\begin{abstract}
This paper derives a general procedure for the numerical solution of the Lindblad equations that govern the coherences arising from multicoloured light interacting with a multilevel system. A systematic approach to finding the conservative and dissipative terms is derived and applied to the laser cooling of gallium. An improved numerical method is developed to solve the time-dependent master equation and results are presented for transient cooling processes. The method is significantly more robust, efficient and accurate than the standard method and can be applied to a broad range of atomic and molecular systems. Radiation pressure forces and the formation of dynamic dark-states are studied in the gallium isotope $^{66}$\Ga.
\end{abstract}

\section{Introduction}

Many recipes \cite{Pillet, Chin, Krems, Julienne, Soldan} now exist in the literature for ultracold molecules starting from a small selection of basic ingredients: trapped, ultracold atoms \cite{Letokhov}. The variety of molecular species is limited by the fact that, to date, only a small number of elements have the requisite electronic structure for direct laser cooling, although sympathetic cooling of molecules with ultracold buffer gases may prove an effective alternative method \cite{Carr}. In order to increase the repertoire of ultracold molecules available, techniques must be developed to cool the large swathes of the periodic table currently inaccessible to laser cooling.

The p-block elements form a particularly important part of the periodic table, but the only such atoms that have been cooled and trapped thus far are the noble gases, paradoxically the least reactive of all elements. There has been, however, some evidence of laser cooling and manipulation of elements within Group 13: a beam of aluminium atoms \cite{Celotta, Lee1995}, has been observed to narrow under laser irradiation and a deceleration force has been demonstrated in gallium \cite{Lee2000} and indium atoms \cite{Metcalf} with recent evidence of sub-Doppler transverse cooling in an indium atomic beam \cite{Meschede2}.  

In general two problems need to be addressed for cooling in the p-block. The first is the tendency of all these elements to absorb light in the UV rather than the visible region when in their ground electronic states. Stable continuous-wave lasers that emit in the UV tend to be complex to operate and generate relatively low power. Sometimes it is possible to find a more convenient, laser-accessible, transition starting from a metastable state, which has been the cooling strategy in the noble gases \cite{Helium1, Helium2} and in the successful Group 13 studies conducted so far. The second is the p-shell itself, with the associated fine structure. The energy splitting due to spin-orbit effects, is many times greater than the hyperfine splitting and necessitates more than one laser (frequency). The usual strategy of repumping can lead to coherent dark states that compress the cooling process \cite{DarkState}.

Group 13 elements fortunately are untroubled by the first problem, which is more an issue with laser technology rather than fundamental physics, having absorption frequencies in the visible and near-UV (with the exception of boron). The $(ns^2, np)$ configuration has a simple doublet ground state,
thus the presence of spin-orbit coupling in the $^{2}{\rm P}$ affords the best opportunity to address the practical issue of cooling in the presence of complex electronic structure. In our first paper \cite{IanGallium}, we conducted time-dependent calculations on the interaction of $^{66}{\rm Ga}$ atoms with counter-propagating laser beams, but the method involved was computationally inefficient, required the derivation of the equations of motion manually and did not easily allow the exploration of optimal experimental parameters, the information experimentalists would find most useful. In this paper, a more sophisticated and elegant method has been developed to quickly and efficiently evaluate possible cooling schemes within these complex structures.

\section{Mathematical Model}

We employ a standard semiclassical treatment throughout, and in this section we briefly introduce the notations, definitions and approximations that we employ in this work.
The notation  is important in forming, in a systematic manner, the 
structure of the differential equations both mathematically and computationally. 

The state of the atom is expressed in a finite set of $N_A$ eigenstates of the hyperfine Hamiltonian ($H^A$) which 
we denote by the single index, $ p \in \{1,2, \dots, N_A \}$. This abbreviation stands for  the ensemble of the hyperfine sub-level quantum 
numbers.  Expanding this notation, we take, $\alpha_p$ to denote the collective label for the fine-structure level and uncoupled 
nuclear spin. That is, $\alpha_p = \{ n_p,L_p,S_p,J_p,I_p\}$ where 
$n_p$ is the electron principal quantum number, $(L_p,S_p)$ are the orbital and spin angular momenta  
 with the resultant coupling $J_p$, and $I_p$ is the  nuclear spin. The basis functions 
 $\vert p \rangle = \vert \alpha_p, F_p, M_p \rangle $ are the orthonormalized  eigenstates with eigenvalues:
 $\varepsilon_p$ (in general not unique).  
 \be
 H^A \vert p \rangle = \varepsilon_p \vert p \rangle 
 \ee
 For convenience we define the corresponding angular frequencies: $\hbar \omega_p \equiv \varepsilon_p$.
 
The  atoms are moving at non-relativistic speeds at all times with the centre-of-mass of the atom   moving with a velocity $\vek{v}$ 
in the laboratory frame, and thus the Galilean transformation can be employed.
The internal state of the atom is expressed by the (variation of constants) expansion:
\be
\vert \Psi (t) \rangle = \sum_{p=1}^{N_A} c^{A}_p(t) \vert p \rangle 
\ee 
Then the density operator, $ \rho \equiv \vert \Psi (t) \rangle \langle \Psi (t) \vert $, has 
the  matrix representation in the time-stationary basis set:
\be
 \rho^A_{pp'} (t) \equiv 
 \langle p \vert \rho^A (t) \vert p' \rangle 
= c^{A}_p(t) c^{A\ *}_{p'}(t)  
\ee

The multicoloured lasers are represented by a superposition of classical monochromatic fields:
\be
\vek{E}(\vek{r}, t) = \sum_{j=1}^{N_L} \vek{E}_j \cos( \vek{k}_j \cdot \vek{r} -\omega_j t+\varphi_j)  
\ee
with the polarizations, wavevectors, freqencies and phases denoted by the usual symbols. 
The superpositions implicitly  allow for any state of polarization. 
Then  the time-varying perturbation can be written:
\be
H^{AL}(t) = - \vek{d} \cdot \vek{E}(\vek{r},t)
\ee
where $\vek{d}$ is the (electronic) dipole-moment operator.

The evolution (Liouville) equation, in the absence of dissipation, takes the form:
\be
{\partial \over \partial t} \rho^A (t) = - (i/\hbar) \left[ H^A+H^{AL}(t), \rho^A (t) \right] 
\ee

The hyperfine Hamiltonian is diagonal in this basis, and in the absence of 
the coupling, $H^{AL}=0$, we have the solution.
\be
\rho^{A} (t) = e^{-iH^A t} \rho^A(0) e^{iH^At } 
\ee 
In other words, the density matrix in the interaction representation is stationary 
with the form: $\rho^A_{pp'}(0) =c^{A}_p(0) c^{A\ *}_{p'}(0)  $.
The dipole approximation is valid for these frequencies, and thus for an electron 
with coordinate $ \vek{r}_e$ with respect to the centre-of-mass of its atom, and thus 
with a position $ \vek{r}+ \vek{r}_e$ in the laboratory frame, we have: 
$ \vek{k}_j \cdot (\vek{r}+ \vek{r}_e) \approx  \vek{k}_j \cdot \vek{r} = \vek{k}_j \cdot \vek{v} t +\phi $.
Then the couplings can be written in the form:
\be
H^{AL}_{pp'}(t) =   \sum_{j=1}^{N_L} \hbar \Omega_{j,pp'}  \cos( \vek{k}_j \cdot \vek{v}t -\omega_j t+\varphi'_j)  
\ee
which reveals the Doppler shift in a simple manner.  The Rabi frequencies determine the selection rules and 
are defined according to:
\be
\Omega_{j,pp'} \equiv     -{1 \over 2\hbar}  \vek{E}_j \cdot \langle p \vert \vek{d} \vert p' \rangle
\ee

It is convenient (analytically) to transform to the interaction representation,
\be
\rho^I_{pp'} \equiv  e^{iH^A t}  \rho^A (t)  e^{i H^A t} 
\ee
and with
$ V(t) \equiv e^{iH^At} H^{AL}(t) e^{-iH^At}$, we have the evolution equation:
\be
{\partial \over \partial t} \rho^I(t) = - (i/\hbar)   \left[V(t), \rho^I(t) \right]    
\ee
At this point it is conventional, though not necessary, to make the 
rotating-wave approximation and discard non-resonant transitions.  That is 
the exponents of the commutator involve the Doppler-shifted detunings 
\be
\delta^{\pm}_{j,pp'} \equiv \omega_p -\omega_{p'} \pm \vek{k}_j \cdot \vek{v}   \mp \omega_j 
\ee
where $\pm$ corresponds to absorption ($+$) or emission ($-$). Once this is done, 
a further unitary transformation to remove the time-dependence (oscillations) from the 
interaction is performed, such that: $H = U(t) V(t) U^{\dag}(t)$. This is only possible when the rotating-wave approximation is enforced, and means that 
the time-integration is now relatively straightforward.

The inclusion of the spontaneous emission terms can be modelled by the Lindblad correction which is written in the form:
\be
{\partial \over \partial t} \rho (t) = - (i/\hbar) \left[ H, \rho (t) \right] + \Gamma \cdot \rho 
\label{eq:master}
\ee
where the dyadic (tensor) for spontaneous emission is denoted by $\Gamma$ and expresses the decoherence 
dynamics of the system.

\section{Outline of computational method}

The master equation (\ref{eq:master}) provides the dynamics of the model, within the rotating-wave approximation.
This equation is used to study the transient and steady-state behaviour of the system. However, naive numerical 
approaches, such as the Euler method,  can lead to unstable integration.  
We propose the following combination of factorisation and modified-Euler method as the integration 
algorithm.  \begin{itemize}

\item{Interaction matrix elements}

The set of atomic hyperfine states $\{ \vert p \rangle \}$ is defined in terms of the quantum numbers, 
and energies. The selection rules for absorption and 
stimulated emission are enforced by the computer program based on 
this data.  Then, for each (allowed) transition, the  multicolour detunings are 
specified and thereby the non-resonant couplings (counter-rotating terms)  identified 
and removed.

\item{Coupling terms}
 
 The reduced matrix elements are assumed as  known, so that one simple uses the 
 Wigner-Eckart theorem \cite{WET} to evaluate the coupling coefficients in terms of 
 the reduced matrix element. 
 
For convenience we introduce the (reduced) Rabi frequency: 
\be
\bar{\Omega}_{j,pp'} \equiv {E_j \over 2\hbar } \vert  \langle p \vert \vert d \vert \vert p' \rangle  \vert
\ee

\item{Dissipation }

The (partial) lifetimes of the excited states are assumed to be known.  That is the 
spontaneous decay process from any upper state to any lower state: 
$\vert p \rangle  \rightarrow \vert p'  \rangle $, is 
given as:

\be
2 \gamma_{pp'}  = {4 \over 3}  
{   \vert \langle \alpha_p F_p \vert \vert d \vert \vert \alpha_{p'} F_{p'}
 \rangle   \vert^2 \vert \omega_p -\omega_{p'} \vert^3 \over (2F_p+1) \hbar c^3} 
\ee

So for each of the dipole allowed transition the relative strength of stimulated and spontaneous emission 
is given by the saturation parameters for each transition:
\be
G_{j,pp'}= 2\bar{\Omega}^2_{j,pp'}/\gamma^2_{pp'} 
\ee

\item{Numerical integration}

Consider \eref{eq:master}. Given the initial state, $\rho(0)$, we step forward in time to $t=h$, using the intermediate helf-step  as follows:
\be
\rho\left( \halfh \right) \approx e^{-i \halfh H} \left[ \rho(0)+ \frac{1}{2}h \Gamma \cdot \rho(0)  \right]  e^{-i \halfh H}
\ee

\be
\rho( h ) \approx e^{-i hH}  \rho(0) e^{-i hH}+ h e^{-i \frac{1}{2}hH(0)}
\left[  \Gamma \cdot \rho(\frac{1}{2}h)  \right]  e^{-i \frac{1}{2}hH}
\ee
The  method is still accurate (only) to first-order in $h$, nonetheless it is much superior in stability to the standard 
Euler method. There is the computational cost of the exponential operators (though these have to be calculated only once within the RWA), but in terms of numerical accuracy 
and efficiency the method is a significant improvement, as will be shown in \sref{Analysis}. The steps are repeated in the usual manner to produce the entire 
evolution over time. The Rabi oscillations are rapid at the beginning followed by slow  damping so that  small time steps 
( $\Omega h \ll 1$ and $\gamma h \ll1 $) are required to ensure that the integration proceeds correctly. However, at  longer times, 
 ($\gamma t \gg 1$)  as the steady state approaches, the characteristic oscillations are slower, and larger time steps can be 
 taken.

\item{Spontaneous emission}

As the spontaneous emission tensor, $\Gamma$, depends on the density matrix, $\rho$, it must be calculated at each time step; this involves solving the equations for the atomic density matrix elements $\rho_{kl} = \rho_{{\alpha_a}{F_a}{M_a} , {\alpha_b}{F_b}{M_b}} = \langle \alpha_a F_a M_a | \rho | \alpha_b F_b M_b \rangle $. To solve the equations each possible pair of ground and excited-state sublevels are considered in turn, and certain selection rules are enforced:
\begin{eqnarray}
\Delta F & = 0, \pm 1 \,({\rm but\,not\,} F=0 \leftrightarrow F'=0) \nonumber \\
\Delta M & = 0\,,\, \pm 1 
\end{eqnarray}
\begin{eqnarray}
M_g-M_g' & = {\rm \,even} \nonumber \\
M_g-M_g' & = M_e-M_e'  
\end{eqnarray}

Each element of the matrix, $\Gamma$, can be described by one of the following four equations \cite{MinoginReview}:

\numparts
\begin{eqnarray}\label{eq:spont1}
\fl \langle{\alpha}_{e_1}{F}_{e_1}{M}_{e_1} |  \,\Gamma \,\rho \,| {\alpha}_{e_2}{F}_{e_2}{M}_{e_2}\rangle  \nonumber \\
 =  -({\gamma}_{{\alpha}_{e_1}{F}_{e_1}}+{\gamma}_{{\alpha}_{e_2}{F}_{e_2}}) \langle{\alpha}_{e_1}{F}_{e_1}{M}_{e_1} | \rho | {\alpha}_{e_2}{F}_{e_2}{M}_{e_2} \rangle \\
\fl\langle{\alpha}_{e}{F}_{e}{M}_{e} |\,\Gamma \,\rho \, | {\alpha}_{g}{F}_{g}{M}_{g}\rangle   =  -{\gamma}_{{\alpha}_{e}{F}_e} \langle{\alpha}_{e}{F}_{e}{M}_{e} | \rho | {\alpha}_{g}{F}_{g}{M}_{g}\rangle \\
\fl \langle{\alpha}_{g_1}{F}_{g_1}{M}_{g_1} | \,\Gamma \,\rho \, | {\alpha}_{g_2}{F}_{g_2}{M}_{g_2}\rangle   \nonumber \\
=  \sum_{{\alpha}_{{e}_1} , {\alpha}_{{e}_2} , {F}_{{e}_1} , {F}_{{e}_2} , {M}_{{e}_1} , {M}_{{e}_2}} ({F}_{g_1}{F}_{g_2}{M}_{g_1}{M}_{g_2} | A | {F}_{e_1}{F}_{e_2}{M}_{e_1}{M}_{e_2}) \nonumber \\
\times \,\langle {\alpha}_{e_2}{F}_{e_2}{M}_{e_2} | \rho | {\alpha}_{e_1}{F}_{e_1}{M}_{e_1} \rangle  \\
\fl \langle{\alpha}_{g} {F}_{g} {M}_{g}' |\,\Gamma \,\rho \, | {\alpha}_{g}{F}_{g}{M}_{g}\rangle \nonumber \\
 \quad =  \sum_{{\alpha}_e  {F}_{e}  {M}_{e} {M}_{e}'} (F_g M_g M_g' | A | F_e M_e M_e') \langle \alpha_e F_e M_e' | \rho | \alpha_e F_e M_e \rangle
\end{eqnarray}
\endnumparts
where
\numparts
\begin{eqnarray}
\fl ({F}_{g_1}{F}_{g_2}{M}_{g_1}{M}_{g_2} | \ A  | {F}_{e_1}{F}_{e_2}{M}_{e_1}{M}_{e_2}) = ({\gamma}_{{\alpha}_{e1} {F}_{e1} , {\alpha}_{g1} {F}_{g1}} + {\gamma}_{{\alpha}_{e2} {F}_{e2} , {\alpha}_{g2} {F}_{g2}}) \nonumber\\
\times \sum_{q=0,\pm1} ({F_{g_1}} {M_{g_1}} 1q | {F_{e_1}} {M_{e_1}})({F_{g_2}} {M_{g_2}} 1q | {F_{e_2}} {M_{e_2}}) \\
\fl ({F}_{g}{M}_{g}{M}_{g}' |\ A  | {F}_{e}{M}_{e}{M}_{e}') \nonumber \\
= 2\gamma_{\alpha_e F_e, \alpha_g F_g} \sum_{q=0,\pm1} ({F_{g}} {M_{g}'} 1q | {F_{e}} {M_{e}'})({F_{g}} {M_{g}} 1q | {F_{e}} {M_{e}})  \\
\fl ({F_{g}} {M_{g}} 1q | {F_{e}} {M_{e}}) = (-1)^{F_g-1+M_e} \sqrt{2F_e + 1} \left( \begin{array}{ccc}
F_g  & 1 & F_e \\ 
M_g & q & -M_e \end{array}\right)  \label{eq:Wigner}
\end{eqnarray}
\endnumparts
The Wigner 3j symbols \eref{eq:Wigner} are calculated using the MATLAB code wigner3j.m \cite{Wigner3j}. 

\end{itemize}

Using this method, the time-evolution and steady-state of the density matrix can be studied. As well as containing information about the populations of each state, the density matrix is also used to calculate the dipole radiation force on the atoms. This force, F, is due to the interaction of the laser field with the induced atomic dipole moment and is given (in one dimension) by;

\begin{equation} \label{eq:forceequation}
F(z,v,t) = -\frac{\partial}{\partial z}U(z,v,t)
\end{equation}
where $U(z,v,t) = -{\rm tr}(\rho\,d)\,E(z,t)$, $E(z,t)$ is the electric field of the lasers along the propagation direction and $d$ is the atomic electric dipole operator (in the atom frame).

\section{Analysis of Method} \label{Analysis}

To analyse the accuracy and stability of our method, it was compared to the basic Euler method \cite{Wallis}. As the exact solution of the density matrix equations for a two-level atom, with ground state, $b$, and excited state, $a$, is well known \cite{Loudon}, this system was used to compare the two methods. Consider an atom at rest ($v=0$), initially with all population in the ground state, $\rho_{aa}=0$. The population of the excited state, $\rho_{aa}$, can be calculated as

\begin{equation} \label{eq:exact}
\rho_{aa} = \frac{G}{(2G+8)} \left\{ 1 - \left[ \cos{\lambda t} + \frac{3\gamma}{2\lambda} \sin{\lambda t} \right] \exp(-\textstyle{3 \over 2}\gamma t) \right\}
\end{equation}
where $G$ is the saturation parameter, $G=2\Omega_{ab}^2 / \gamma^2$, with $\gamma$ the spontaneous decay rate and $\lambda = {\textstyle \gamma\over 2} \sqrt{2G - 1}$.

\begin{figure}[h] 
\centering
\includegraphics[width=0.6\linewidth]{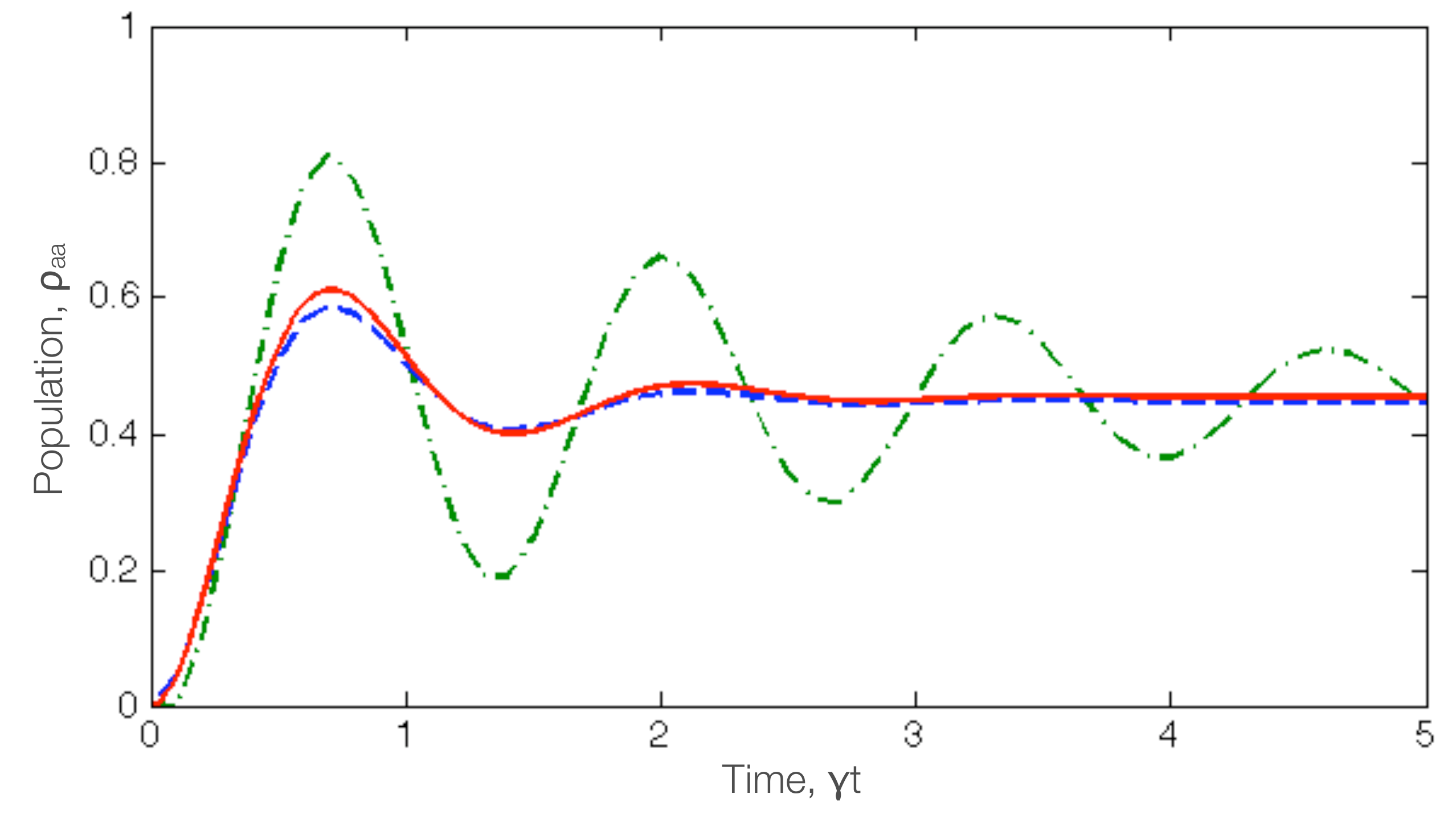}
\caption{Population $\rho_{aa}$ for the excited state of two-level atom, in which $\rho_{aa}(0)=0$, $G=10$, $\delta=0$, $\gamma=1$, $h=0.1$, at velocity $kv/\gamma=0$. Figure shows the results of the numerical integration method presented in this work (dashed), compared to the Euler method (dot-dashed) and the exact solution (solid).}\label{fig:eulercomparison}
\end{figure}

\Fref{fig:eulercomparison} shows the time-evolution of the excited-state population $\rho_{aa}$, calculated on resonance ($kv=\delta=0$) by both the Euler method and the method presented in this work, compared with the exact solution given by  \eref{eq:exact}. It is clear that even for this simple system the new method converges significantly more rapidly than the Euler method. This slow convergence of the Euler method becomes even more marked as $\gamma$ is reduced. At $\gamma=0$ (pure Rabi oscillation), and in calculations far off-resonance, the Euler method fails to converge unless step sizes are extremely small ($h<$0.001) whereas the method presented is highly stable, even at step sizes over 100 times larger. 

Comparisons for more complex systems were not possible as the extremely small step-sizes needed to counteract the instability of the Euler method meant calculations could not be carried out in realistic timescales. In contrast, the method presented here remained stable over a wide range of detunings and step-sizes. It is clear therefore that the new method provides a marked improvement on accuracy, stability and computational cost.

Previous calculations \cite{IanGallium} on a similar system to that outlined in this paper depended on the construction, by hand, of a set of $n\times n$ density matrix equations, where $n$ is the number of individual magnetic sublevels in the cooling scheme. For small systems this is a relatively simple task but as the number of magnetic sublevels increases, so does the number of equations needed. Soon it becomes hugely time-consuming to construct and solve these equations, with the potential for human error also increasing. The most difficult part of the master equation \eref{eq:master} to calculate is the section containing the spontaneous emission terms, $\Gamma(\rho)$, which often has to take into account many different decay channels, alongside their associated Clebsch-Gordan coefficients. The benefit of the method presented here is that, with the only input being a small number of quantum numbers, these terms are calculated automatically within the numerical code, and placed in a matrix. This abolishes the need for time-consuming construction of equations or matrices by hand.

\section{Application to a $\Lambda$-system}

\begin{figure}[h] 
\centering
\includegraphics[width=0.5\linewidth]{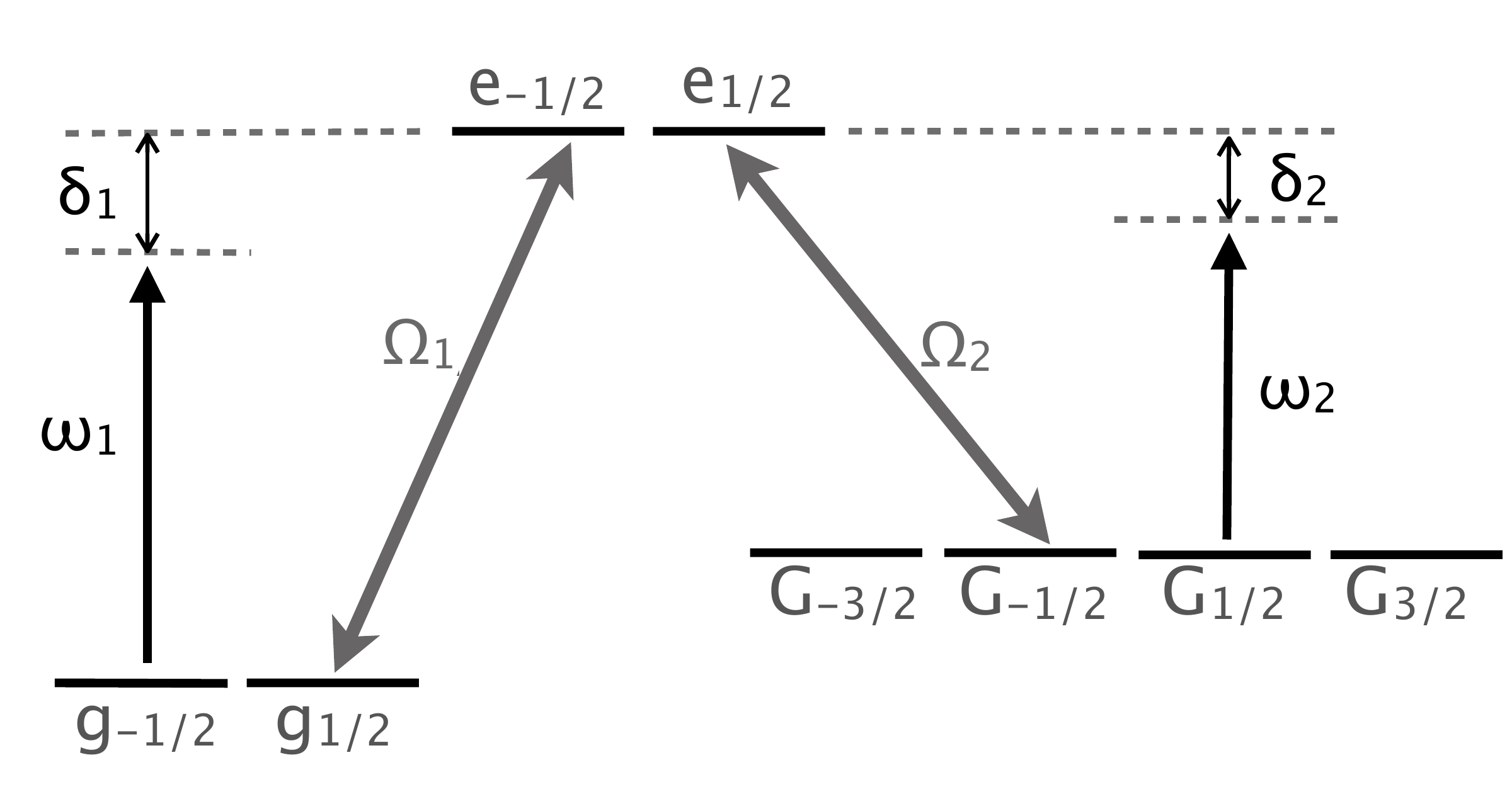}
\caption{Schematic energy level diagram for 2-(4)-2 multilevel atom, corresponding to the $^2\Pterm_{1/2}$, $^2\Pterm_{3/2} \leftrightarrow \, ^2\Sterm_{1/2}$ transition in Group 13 atoms with nuclear spin I=0. $\Omega$ corresponds to the Rabi frequency, $\omega$ is the laser frequency and $\delta$ is the laser detuning.}\label{fig:242energylevel}
\end{figure}

To show the possibilities of this new method, in this paper the method presented is applied to a simple multi-level system. Consider now a $\Lambda$-system involving the lower manifolds $|\g\rangle$, $|\G\rangle$, and the excited state $|\e\rangle$, as shown in \fref{fig:242energylevel}. This system is assumed to correspond to the $^2\Pterm \to \,^2\Sterm$ transition in a Group 13 atom with nuclear spin I=0, specifically $^{66}\Ga$, a beta emitter and 
putative radiopharmaceutical for PET imaging which can be produced with a 
biomedical cyclotron \cite{Ga66}. This scheme could also be applied to a transition to the first hyperfine excited level of a Group 1 atom with nuclear spin I=1 (eg $^6$Li).

For the example presented, two laser frequencies are considered, $\omega_1$ and $\omega_2$, red-detuned by $\delta_1$ and $\delta_2$ respectively. One laser beam corresponds to the $F_g = 1/2 \to F_e = 1/2$ transition, with the other used to pump $F_G=3/2 \to F_e=1/2$. The scheme considered here assumes the use of counter-propagating circularly-polarized laser fields, with a $\sigma^+$-polarized beam driving the $ M_F \to  M_F +1$ transition, and a $\sigma^-$ laser driving $ M_F  \to  M_F -1$. 

For simplicity it is assumed that the radiative widths for the two transitions are approximately equal, $\gamma_1\approx\gamma_2=\gamma$, that $k_1\approx k_2 = k$, and that the saturation parameters, $G_1=G_2=1$. Tests with a range of parameters show that these assumptions have little effect on the results of the calculations.

\subsection{Population Distribution}

\begin{figure}[h!] 
\centering
\includegraphics[width=0.6\linewidth]{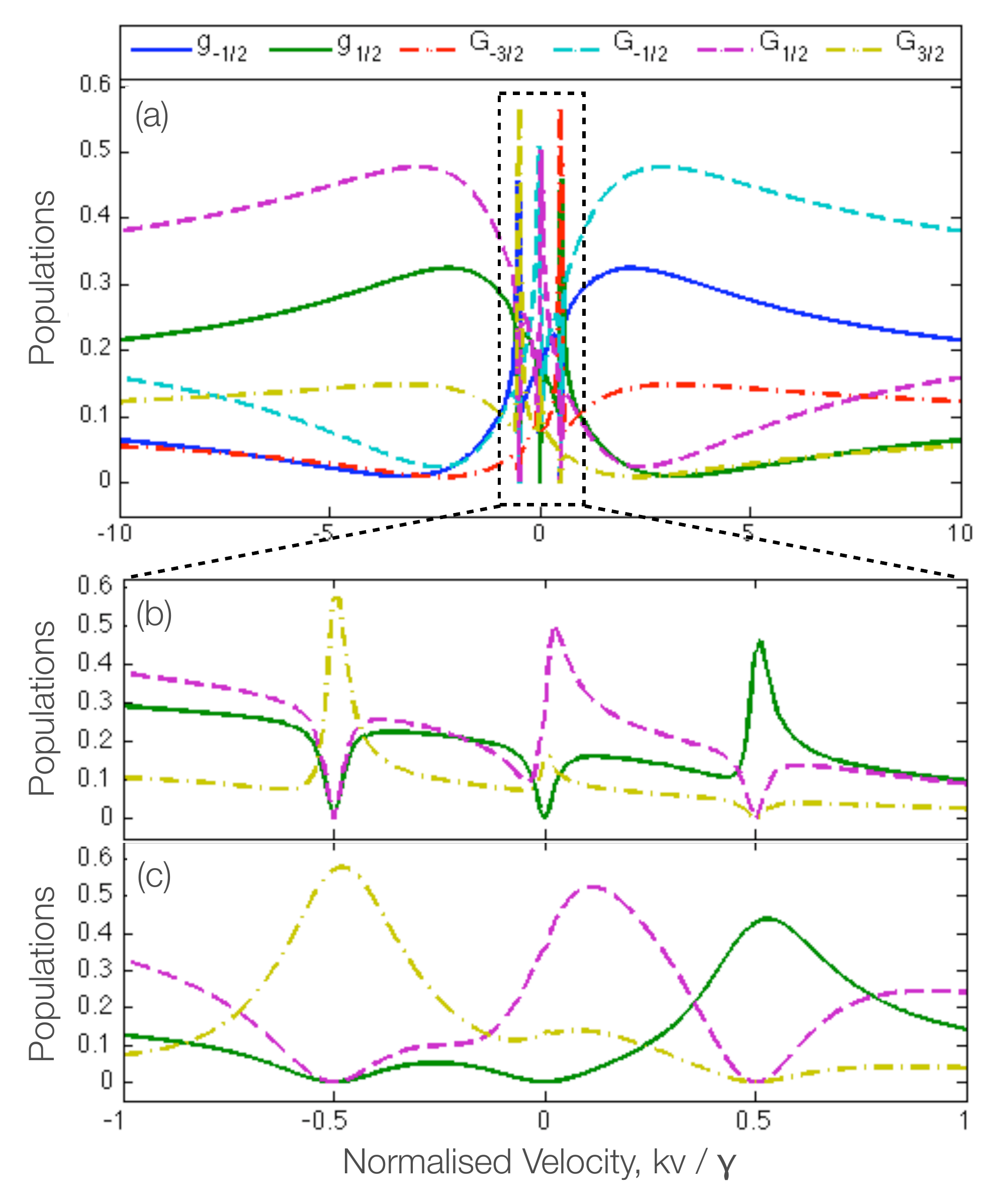}
\caption{Ground-state populations g$_{\pm1/2}$ (solid), G$_{\pm1/2}$ (dashed) and G$_{\pm3/2}$ (dot-dashed) of a 2-(4)-2 multilevel atom (\fref{fig:242energylevel}) in the steady state as a function of velocity $v=v_z$. Radiative widths are equal, $\gamma_1=\gamma_2=1$, as are laser intensities, $G_1=G_2=1$, while detunings are $\delta_1=-3\gamma$, $\delta_2=-2\gamma$. The figure shows sharp two-photon resonances at velocities $kv/\gamma=\pm 0.5$ and $kv/\gamma=0$. Figure (b) shows an enlarged view of (a), with $G=1$, but as the population distribution is highly symmetrical the states g$_{-1/2}$, G$_{-1/2}$ and G$_{-3/2}$ are omitted from (b) for clarity. (c) also shows an enlarged view, in this case with $G=10$, and at this higher laser intensity broadening of the two-photon population peaks is observed. }\label{fig:steadystate_montage}
\end{figure}

The population distribution of the system is investigated over a range of atomic velocities, along with the dipole radiation force, for both the time-dependent and steady state cases. The steady state result is obtained numerically, for any initial conditions, by time-dependently solving the master equation at very long timescales, until the populations converge. \Fref{fig:steadystate_montage}(a) shows the  steady-state ground-state population distribution for a 2-(4)-2 multilevel atom, as shown in \fref{fig:242energylevel}, with laser detunings $\delta_1=-3\gamma$ and $\delta_2=-2\gamma$. 

In \fref{fig:steadystate_montage}(a), it can be seen that as the velocity of the atom nears resonance velocities ($kv/\gamma = \pm 2$, $kv/\gamma=\pm 3$) there is a transfer in populations giving rise to a broad peak, as expected. The system possesses a left-right symmetry, with a change in the sign of the velocity being equivalent to a change in the sign of the quantum number $ M_F $. Sharp population structures are also observed closer to zero velocity. These resonances are due to two-photon processes producing ground-state coherences which vary rapidly at low velocities, leading to sharp variations in ground-state populations. The positions of these resonances can be predicted simply from the energy conservation law \cite{TwoPhoton}. Two-photon transitions within the ground state $|\G\rangle$ do not change the energy of the atom, so $(\omega_2\pm kv)-(\omega_2\mp  kv)=0$ and the resonance occurs around zero velocity. However, in the case of two-photon transitions between states $|\g\rangle$ and $|\G\rangle$ the energy of the atom is changed;
\begin{eqnarray}
(\omega_1\pm kv)-(\omega_2\mp kv)=(\omega_1+\delta_1)-(\omega_2+\delta_2)
\end{eqnarray} 
where $\omega_i$ is the laser frequency used and $\delta_i$ is the detuning (as shown in \Fref{fig:242energylevel}). These resonances, therefore, occur at velocities $kv=\pm(\delta_1-\delta_2)/2$, in this case at $kv=\pm0.5\gamma$. \Fref{fig:steadystate_montage}(b) gives an expanded view of the region in which these population shifts occurs, and as \fref{fig:steadystate_montage}(c) shows, increasing the laser intensity leads to a broadening of these peaks. Conversely, a low saturation parameter gives rise to much narrower resonance structures. As the population distribution is highly symmetric, for clarity only the levels ${\rm g}_{1/2}$, ${\rm G}_{1/2}$ and ${\rm G}_{3/2}$ are shown in the expanded region.

\begin{figure}[h] 
\centering
\includegraphics[width=0.6\linewidth]{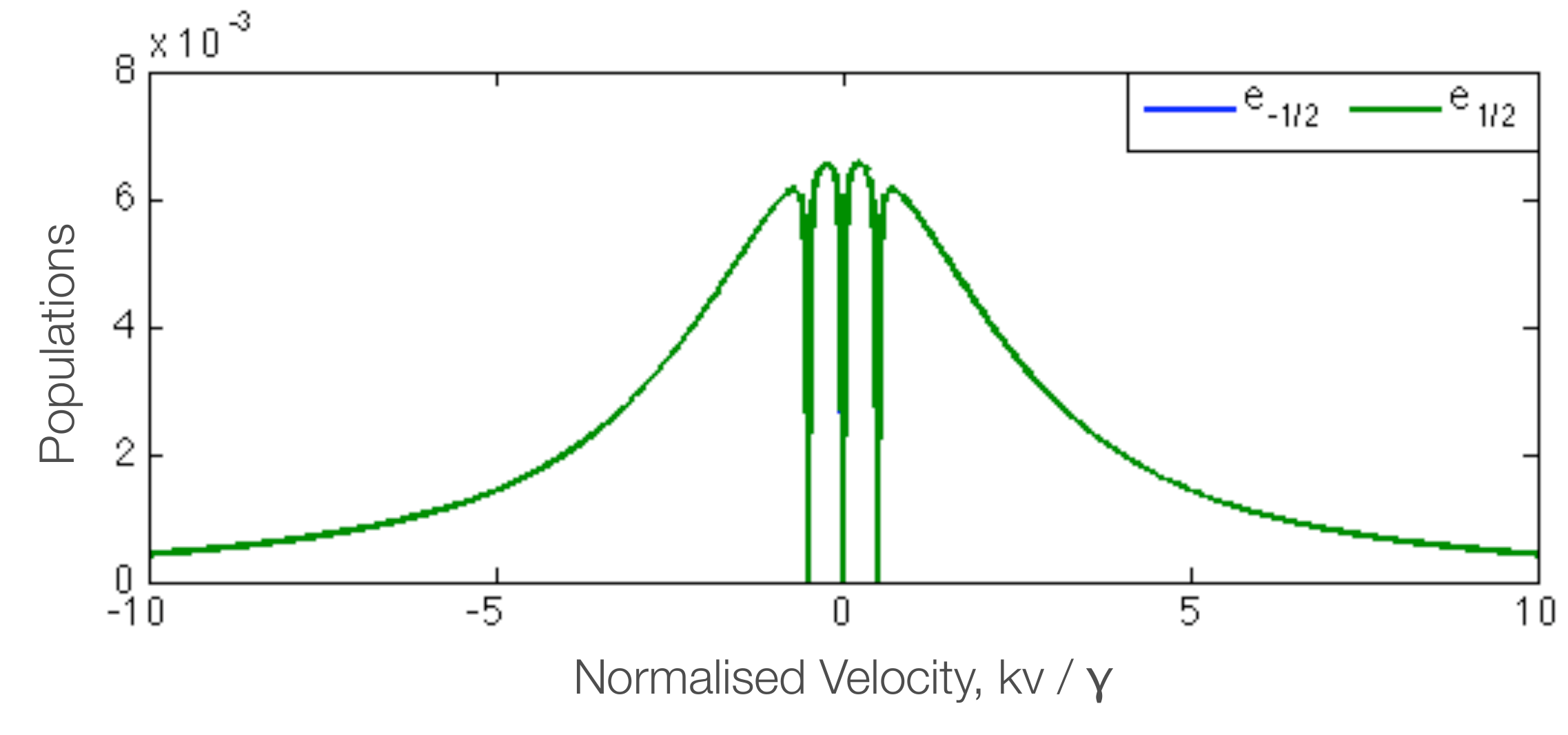}
\caption{Steady-state excited populations as a function of velocity $v=v_z$. The populations for each of the excited states is identical, with sharp drops to zero observed as a result of two-photon resonance processes. These demonstrate destructive interference occurring with the formation of dark states; saturation parameter $G=1$, detunings $\delta_1=-3\gamma$, $\delta_2=-2\gamma$.}\label{fig:steadystateexcited}
\end{figure}

Steady-state excited populations are shown in \fref{fig:steadystateexcited}; due to symmetry, the population distribution for each excited-state sublevel is identical. Ground-state coherences caused by two-photon processes result in the populations falling sharply to zero at these resonance points. The presence of these structures at all laser intensities clearly demonstrates the destructive atomic interference occurring in the system, with the formation of dark states.

\begin{figure}[h] 
\centering
\includegraphics[width=0.6\linewidth]{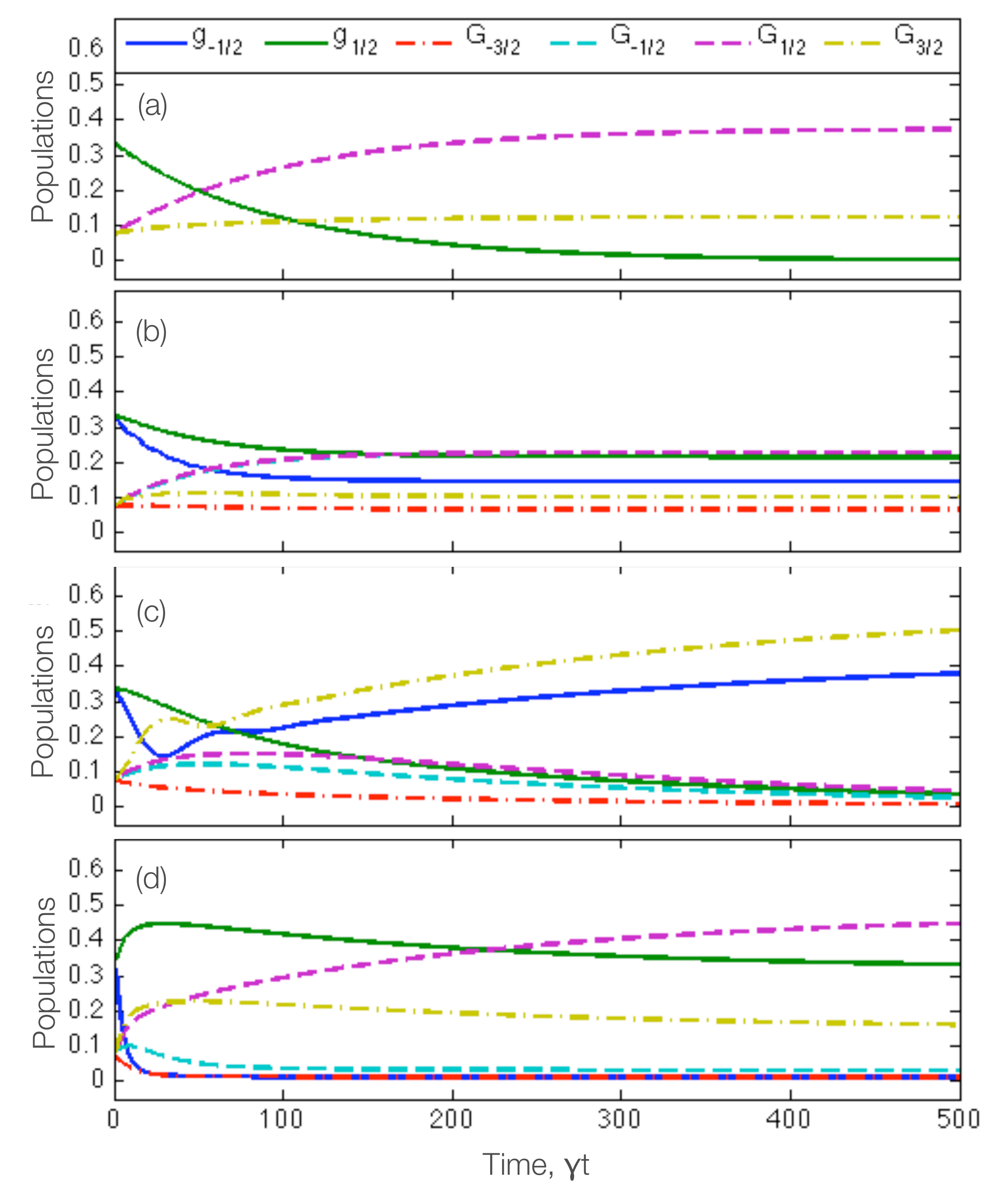}
\caption{Evolution of ground-state populations g$_{-1/2}$, g$_{1/2}$ (solid), G$_{-1/2}$, G$_{1/2}$ (dashed) and G$_{-3/2}$, G$_{3/2}$ (dot-dashed) at (a) $kv/\gamma =0$, (b) $kv/\gamma =-0.25$, (c) $kv/\gamma=-0.5$ (two-photon resonance velocity) and (d) $kv/\gamma=-3$ (at a resonance velocity). At $kv/\gamma=0$, (a), the symmetry of the system dictates that sublevels with the same modulus $|F,  M_F |$ have identical populations. This symmetry is broken for non-zero velocities, as shown in (b). At two-photon resonance velocities, as illustrated by (c) $kv/\gamma=-0.5$, markedly different behaviour is observed, particularly in the $g_{-1/2}$ population, which rises to a maximum but falls in every other case. Population oscillations also occur. In subfigure (d), taken at a resonance velocity $kv=\delta_1$, rapid changes in population are observed at short timescales ($\gamma \timeT < 20$). The state $\g_{-1/2}$ is depopulated quickly, with a resulting gain in population for the $\G_{-1/2}$ and $\G_{3/2}$ sublevels. Similarly, population in the $\G_{-3/2}$ state decreases, resulting in a gain for the $\g_{1/2}$ sublevel. After this time the populations approach their  steady-state solution, with states with $ M_F >0$ most highly populated: saturation parameter $G=1$, detunings $\delta_1=-3\gamma$, $\delta_2=-2\gamma$, initial ground-state populations: $\rho_{gg}=0.344$, $\rho_{GG}=0.078$}\label{fig:popevolmontage}
\end{figure}

We now consider the formation of these resonances and dark states by studying the time-dependent evolution of the populations. \Fref{fig:popevolmontage} shows how the ground-state population distribution varies with time at different velocities. In each of the figures the initial conditions are taken as the Boltzmann distribution in a gas of gallium atoms at 1500K, the typical temperature of a gallium effusive cell. Initially, $34.4\%$ of the atoms are in each of the $^2\Pterm_{1/2}$ states, with $7.8\%$ in each of the $^2\Pterm_{3/2}$ sublevels. At $kv/\gamma=0$, (a), a symmetry is observed as expected, resulting in three sets of overlapping populations. Due to the two-photon processes and the formation of dark states within the state $|\G\rangle$, the population in the $|\g\rangle$ state decays to zero. As the velocity decreases slightly to $kv/\gamma=-0.25$ (b), the six different populations can be seen to reach their  steady-state distribution quickly. As the velocity is now non-zero, the symmetry of the system is broken, and the behaviour of individual populations can be observed. \Fref{fig:popevolmontage}(c) shows the population evolution at $kv/\gamma=-0.5$, one of two-photon resonances between the two fine structure states. In this figure oscillations are observed, before the populations of states G$_{3/2}$ and g$_{-1/2}$ rise to a maximum. All other populations tend to zero. This behaviour is notably different from that at other velocities. In subfigure (d), the internal processes can be seen clearly. At first ($\gamma \timeT < 20$) population is transferred rapidly from the $\g_{-1/2}$ state, resulting in a gain in population, via the excited state $\e_{-1/2}$, for the $\G_{3/2}$ and $\G_{-1/2}$ sublevels. Similarly, population in the $\G_{-3/2}$ state decreases, resulting in a gain for the $\g_{1/2}$ sublevel. After this time, the populations approach their  steady-state distribution, with the $\g_{1/2}$, $\G_{1/2}$ and $\G_{3/2}$ states populated while the other three ground-state populations tend to zero.

\subsection{Dipole Radiation Force}

\begin{figure}[h] 
\centering
\includegraphics[width=0.65\linewidth]{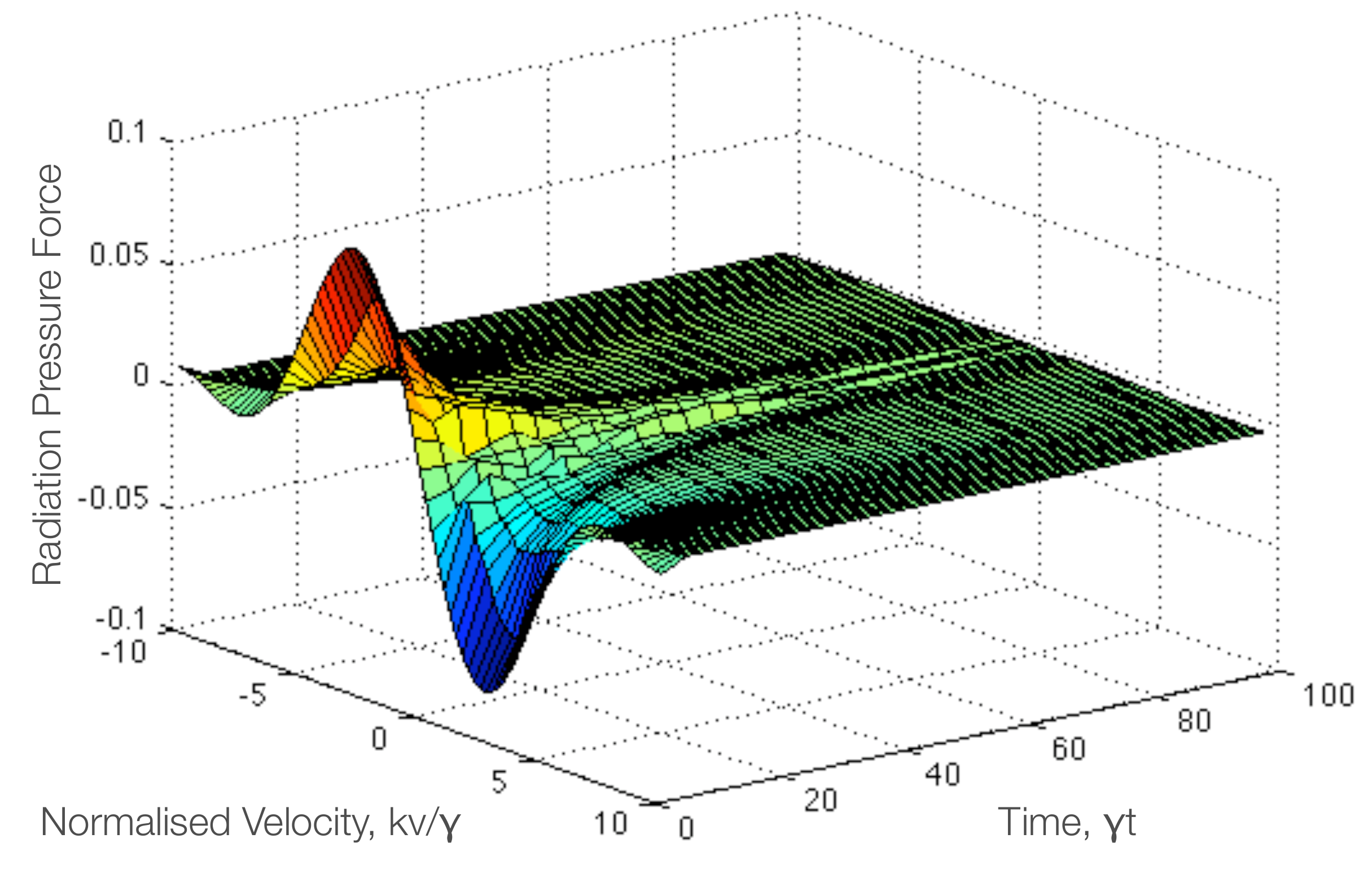}
\caption{Time dependent dipole radiation force for the 2-(4)-2 multilevel atom (\fref{fig:242energylevel}) as a function of normalized velocity. Saturation parameter $G=1$ and detunings $\delta_1=-3\gamma$, $\delta_2=-2\gamma$.}\label{fig:G1forceevol}
\end{figure}

The radiation pressure force on an atom, using \eref{eq:forceequation}, can be easily calculated from the atomic density matrix, by summing the force contributed by each individual coherence. For the 2-(4)-2 level system, this is found using the following equation;
\begin{eqnarray}
\fl F =  2\hbar k \Omega {\rm Im} \Bigg( \frac{1}{\sqrt{3}}\,\rho\,_{g_{-{1\over2}} e_{1\over2}} - \frac{1}{\sqrt{3}}\,\rho\,_{g_{1\over2} e_{-{1\over2}}} + \frac{1}{2}\, \rho\,_{G_{-{3\over2}} e_{-{1\over2}}} + \frac{1}{2\sqrt{3}}\, \rho\,_{G_{-{1\over2}} e_{1\over2}} \nonumber \\
 - \frac{1}{2\sqrt{3}}\, \rho\,_{G_{1\over2} e_{-{1\over2}}} -\frac{1}{2} \,\rho\,_{G_{3\over2} e_{1\over2}}
\Bigg)
\end{eqnarray}
Calculations show that this system does not possess a cooling force in the steady state. The absence of force is a result of coherent population trapping, leading to formation of a velocity-independent dark state which inhibits the cooling cycle. However, time dependent calculations show that a transient force is present, as seen in \fref{fig:G1forceevol}, and decays to zero as the steady state is approached. Although the ground-state populations are found to continue evolving past $\gamma\timeT=500$, this transient force has decayed to zero by around $\gamma\timeT=40$, corresponding to the peak in $\g_{1/2}$ and $\G_{3/2}$ populations in \fref{fig:popevolmontage}(d). The initial force observed (at $\gamma\timeT = 1$) is found to continue oscillating at larger velocities, with a regular frequency but low amplitude, before dying away. 

\newpage
\section{Optimal parameters}

Analysis of individual coherences in the atomic density matrix shows that, for this system, the majority of the force is contributed by the $^2{\rm P}_{1/2} \to ^2{\rm S}_{1/2}$ transition. Calculations were also carried out to simulate cooling if all population is transferred to the $^2{\rm P}_{3/2}$ state, for example by a STIRAP process. These conditions led to a much reduced initial force. This indicates that to maximize the force experienced by the atoms it is preferable to have as large a proportion of the population as possible in the $^2{\rm P}_{1/2}$ state. This would need to be taken into consideration in any experimental proposal.

\begin{figure}[h!] 
\centering
\includegraphics[width=0.8\linewidth]{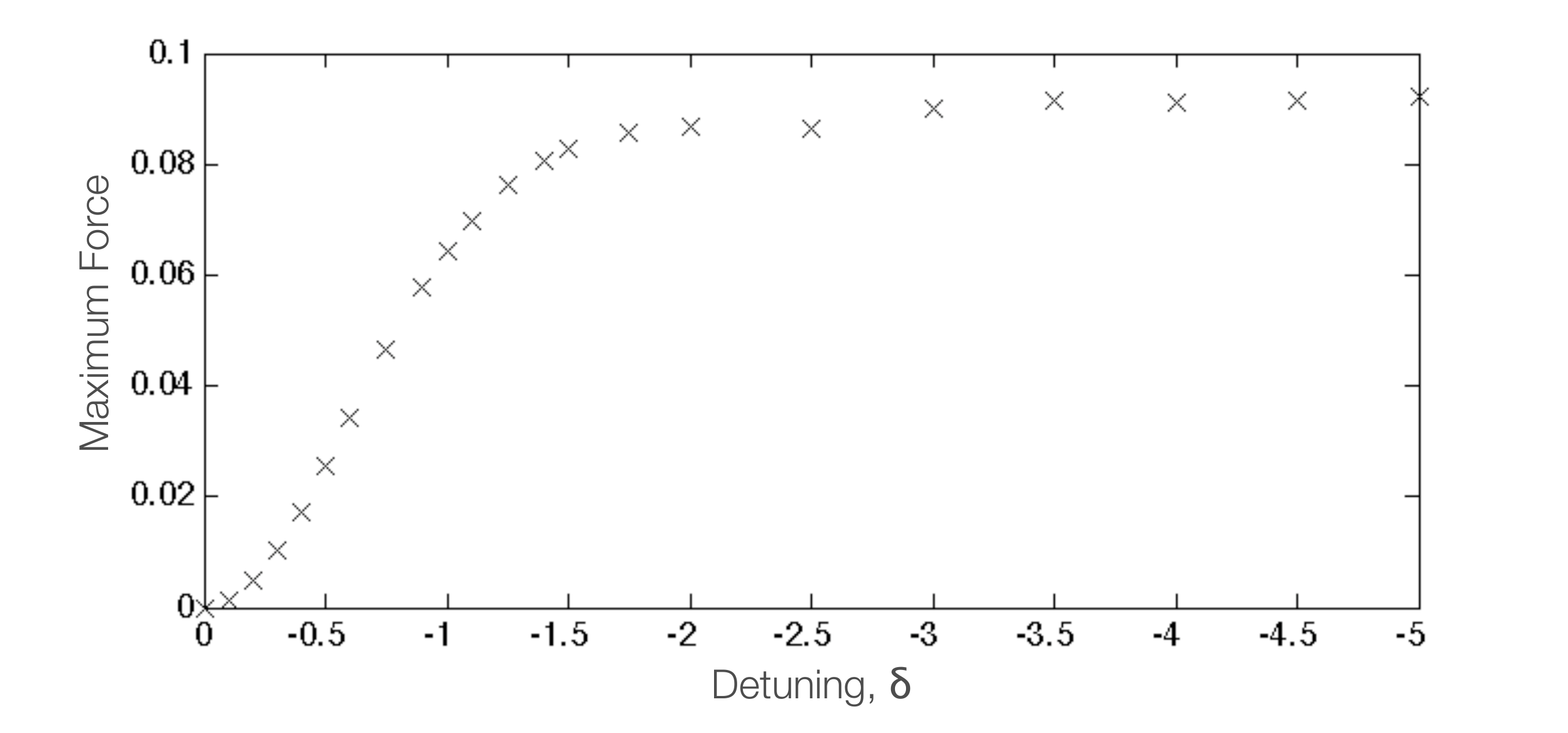}
\caption{Variation of the maximum radiation force with laser detuning, at resonance velocity, $kv=\delta$; Saturation parameter $G=1$, detuning $\delta=\delta_1=\delta_2$}\label{fig:Detuning_vs_Force}
\end{figure}

Laser cooling schemes provide an extremely large parameter space, and as an example, we have investigated the laser detuning, $\delta$. \Fref{fig:Detuning_vs_Force} shows how the maximum initial force experienced by the atoms varies with laser detuning. The force is measured at resonant velocity, $kv=\delta_1=\delta_2$. It can clearly be seen that increasing the detuning above $-2\gamma$ does not result in a significant increase in force. For symmetry reasons the force decays to zero at $kv=\delta=0$.

It has also been observed that a small detuning difference between the two lasers is beneficial. The force is found to decay more slowly when the two detunings vary by a small amount, approximately when $(\delta_1-\delta_2)=-0.5\gamma$. It is possible that this small detuning difference minimizes interference effects between the two laser beams.

\section{Conclusion and Outlook}

\begin{figure}[h] 
\centering
\includegraphics[width=0.9\linewidth]{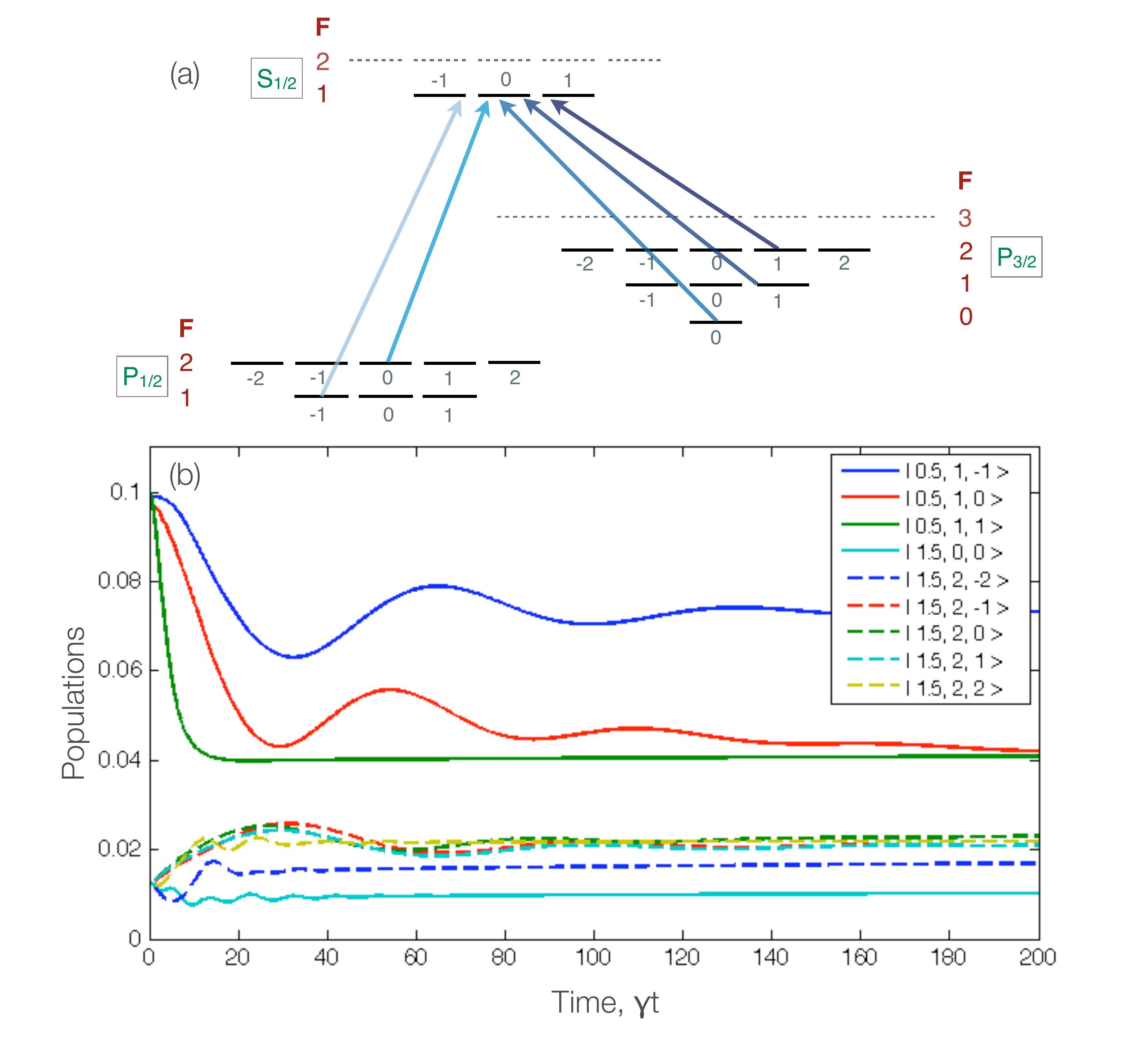}
\caption{(a) Energy level schematic for $^{69}\Ga$. Transitions forming a possible 3-5-(1-3-5)-3 five-colour cooling cycle are highlighted. (b) Example of the evolution of selected $^2\Pterm_{1/2}$, $^2\Pterm_{3/2}$ ground-state populations in $^{69}$Ga, at resonance velocity $kv=\delta$. Assumptions: all laser saturation parameters $G_i=1$, all radiative widths are equal $\gamma_i=1$, and all detunings $\delta_i=-2\gamma_i$. For clarity only a selection of states are shown, and are labelled according to quantum numbers $| J, F,  M_F  \rangle$. Initial populations are distributed equally amongst magnetic sublevels in each fine structure ground state, with $80\%$ in the $^2\Pterm_{1/2}$ state, and $20\%$ in the $^2\Pterm_{3/2}$.}\label{fig:Ga69PopEvol}
\end{figure}

An improved method for calculating the transverse cooling of atoms with complex ground-state structures has been presented, and the optimal detunings determined to maximize the force on an atom such as $^{66}$Ga. The method is efficient and can be adapted to considerably larger multilevel systems. As an example of the kind of complex system that can be considered, the $\Lambda$-shaped cooling cycle outlined above is applied to a Group 13 isotope with nuclear spin $I=3/2$, for example $^{69}\Ga$. The ground state and first excited state contain 32 magnetic sublevels, requiring 1024 density matrix equations for a full treatment. Using the method presented this calculation requires the formation of only three $32\times32$ matrices. One suggested cooling cycle is a 3-5-(1-3-5)-3 scheme, involving only one of the hyperfine excited states. This requires five different coloured lasers, as shown in \fref{fig:Ga69PopEvol}(a). The evolution of ground-state populations at resonance velocity $kv=\delta$ are shown in \fref{fig:Ga69PopEvol}(b). For clarity only a selection of states are shown, and are labelled according to quantum numbers $| J, F,  M_F  \rangle$.

An interesting atomic system for future investigation is that found in the stable isotopes of Group 13 atom thallium, $^{203}$Tl and $^{205}$Tl, which both have $I=1/2$ and are important candidates for sensitive measurements of parity non-conservation \cite{PNC}.

In principle this method could also be used to model the laser cooling of molecular systems in which a suitable cooling scheme has been identified. The simplicity of the method means that electric or magnetic fields can be incorporated into the Hamiltonian matrix with little difficulty, as can time-dependent pulsed or shaped laser fields, as well as complex repumping schemes. This could be of great use in investigating methods for overcoming dark state resonances which hamper the laser cooling process.

\ack
LR would like to thank the Department for Employment and Learning (NI) for funding this PhD research.

\end{document}